\documentclass[preprint,review,12pt]{elsarticle}

\usepackage{graphicx}
\usepackage{amssymb,amsmath,mathrsfs,amsfonts}
\usepackage{epstopdf}
\usepackage{algorithm}
\usepackage{algorithmic}
\usepackage{subcaption}
\usepackage{float}
\usepackage{nccmath}
\usepackage{lineno,hyperref}
 \usepackage{setspace}
 
 \newtheorem{theorem}{Theorem}
 
 \newdefinition{rmk}{Remark}
 \newproof{proof}{Proof}
 \newproof{pot}{Proof of Theorem \ref{thm2}}
  \newtheorem{assumption}{Assumption}
  \newtheorem{remark}{Remark}

  \usepackage{lipsum}
\usepackage{mathtools}
\usepackage{cuted}
\usepackage{comment}

\usepackage{lineno}

\journal{To be decided}

\begin{document}

\begin{frontmatter}




\title{ A Variable Power Surface Error Function backstepping based Dynamic Surface Control of Non-Lower Triangular Nonlinear Systems }

\author[a]{Abdulrazaq Nafiu Abubakar} 
\author[a,b,c]{Ali Nasir*}
\author[a]{Md Muzakkir Quamar} 
\address[a]{Control and Instrumentation Engineering Department, KFUPM, Dhahran 31261, SAUDI Arabia}
\address[b]{Interdisciplinary Research Center for Intelligent Manufacturing and Robotics, KFUPM, Dhahran 31261, SAUDI Arabia}
\address[c]{Interdisciplinary Research Center for Aviation and Space Exploration (guest affiliate), KFUPM, Dhahran 31261, SAUDI Arabia

*ali.nasir@kfupm.edu.sa}











\begin{abstract}
A control design for error reduction in the tracking control for a class of non-lower triangular nonlinear systems is presented by combining techniques of Variable Power Surface Error Function (VPSEF), backstepping, and dynamic surface control. At each step of design, a surface error is obtained, and based on its magnitude, the VPSEF technique decides the surface error to be used. Thus, the backstepping-based virtual and actual control law is designed to stabilize the corresponding subsystem. To address the issue of circular structure, a first-order low-pass filter is used to handle the virtual control signal at each intermediate stage of the recursive design. The stability analysis of the closed-loop system demonstrates that all signals indicate semi-global uniform ultimate boundedness. Moreover, by using the switching strategy of the control input using the VPSEF technique suitably, it is possible to ensure that the steady-state tracking error converges to a neighborhood of zero with an arbitrarily very small size. The effectiveness of the proposed concept has been verified using two different simulated 
demonstrations.

\end{abstract}


\end{frontmatter}

\section{Introduction}
The backstepping method is a frequently used strategy for designing controllers and analyzing stability in nonlinear systems that have a strictly triangular structure \cite{maaruf2023hybrid}, \cite{zhou2007adaptive}, \cite{maaruf2022survey}. Nevertheless, the system architecture of triangular nonlinear systems may be influenced by many uncertainties, such as modeling inaccuracies, unidentified parameters, time delays, disturbances, unknown faults, input and state delays, and so on \cite{marani2023non},\cite{li2022stabilization},\cite{krishnamurthy2017global},\cite{krishnamurthy2011singular},\cite{cai2022semi},\cite{li2023optimized}. Several design approaches for adaptive control were developed specifically for lower triangular nonlinear systems with linearly parameterized uncertainty using backstepping techniques \cite{merei2023adaptive},\cite{umutlu2024adaptive},\cite{cai2022decentralized}. while online approximation methods based adaptive backstepping control approaches were presented for specific types of lower triangular systems with nonlinearly specified uncertainty for example adaptive control based neural network \cite{liu2022event}, Adaptive fuzzy control \cite{xu2023adaptive}. Traditional backstepping based control design approaches have a limitation related to the complexity of controller design. In order to simplify this control design, several approaches such as dynamic surface control (DSC) are used \cite{ xing2023finite}, \cite{bi2022event}, \cite{yao2023backstepping}, minimal learning parameter \cite{li2009dsc}, \cite{yang2004combined}, \cite{yang2003adaptive}, online single neural network approximation \cite{wan2022neural}, \cite{ma2017adaptive} and so on were integrated into the framework of backstepping-based control design.

Although there has been much research on the management of lower triangular nonlinear systems using backstepping design, there are comparatively few findings in research about the control of nonlinear systems having non-lower triangular structures. A significant challenge in designing controls for nonlinear systems that are not lower triangular is the circular structure of the controller. This circular structure arises from the structural characteristics of the class of nonlinear systems and the differentiation of virtual control laws in the backstepping design stage. Indeed, in certain lower triangular nonlinear systems with pure-feedback structure, the issue of circular structure also arises in backstepping based control design. Most recently relevant research to the backstepping approach as the circularity problem solver include \cite{wang2019backstepping} which introduces a tracking control design technique suited for nonlinear systems with good synthesis and filtering to deal with slow rates and circular structure drawbacks. The authors in \cite{bai2016backstepping} enhance the capacity to control trajectories for spherical rolling robots by using the backstepping approach on a ball-pendulum system. The performance of the controller is shown via simulations. the study in \cite{niu2022backstepping} examine output feedback approaches for integrators that are globally stable and systems that include uncertainties and also consider the use of delayed static observers and adding backstepping techniques to expand the range of control applications. A control design strategy based on adaptive backstepping was presented in \cite{wang2011neural} for a specific kind of uncertain affine pure-feedback nonlinear systems. In the study, a DSC approach was used to resolve the issue of circular construction in the design. Nevertheless, the dynamic surface control reduces the circular structure, but there are tracking errors that tends to distabilize the system at initial stage with high overshoot which is due to the initial magnitude of the error, thus, reducing the performance of the controller. These errors tends to reduce as the controller becomes more adaptive. In this situation, the control performance is good but the error changes with time. Therefore, there is a need to control  the magnitude of the error. 

In this work, a problem of tracking error reduction for non lower triangular nonlinear class of system is considered. A Variable Power Surface Error Function (VPSEF) backstepping based DSC design method will be developed for an $n$th-order non lower triangular nonlinear system. At each step of the control design, say the intermediate step $i$ of the design being recursive, a switching full state feedback virtual control law is firstly designed to stabilize the corresponding switched subsystem $i=1,2,3,...n-1$. This control law will switched based on the magnitude of the error i.e. being higher or lower error at a time $t$. Therefore, this entails that, the surface error of the control law will be different in respect of the magnitude of the error. By using a first-order low-pass filter to process the switching virtual control signal and allowing the filtered output to be used in the next design stage, the issue of circular construction may be efficiently resolved. Finally, in the last step $n$, a complete state feedback real switching control is provided to stabilize the final subsystem. To demonstrate reliability of the suggested method, the stability analysis of the system under study is shown in a subsection below. The closed-loop system signals are uniformly finally bounded, indicating that the system tracking error may converge to a small neighborhood of zero by carefully reducing the error through the switched strategy and selecting parameters.

Motivated by the literature and the discussion above, this article studies the error reduction in trajectory tracking of a special class of nonlinear system. First, a switching virtual controller is designed for the $i$th subsystem considering the surface error of the backstepping technique to switched based on the magnitude of the error at an instant time $t$. We then apply the first order low pass filter to process the virtual switching control signal and finally the full state feedback actual switching control is designed to stabilize the last subsystem.  To the best of our knowledge, this is the first time the Switching control of DSC based backstepping, is proposed for the tracking control of a non lower triangular nonlinear system. The main contributions of this paper are enumerated as follows:

\begin{enumerate}
    \item Contrary to the the backstepping methods in  \cite{wen2024optimized}, \cite{wang2019backstepping}, \cite{cai2020adaptive}, \cite{xing2023finite}, we used the switching control strategy to eliminate the effect of high and low surface errors related to the tracking performance using the VPSEF technique.
    
    \item Compared with the control methods in \cite{wang2019backstepping}, \cite{hou2023adaptive} and \cite{cai2020adaptive} the switched VPSEF  backstepping based DSC show a superior trajectory tracking in respect of the tracking error and change in reference signals.
    
    \item The Control strategy proposed guarantees stability, fast and finite time convergence to the desired signals.
\end{enumerate}

The paper is outlined as follows. The nonlinear dynamic model of the non lower triangular nonlinear system is given in Section \ref{sec2}. The proposed switching VPSEF backstepping based DSC are designed in Section \ref{sec3}. Stability analysis is proved in \ref{sec4}. Simulation results are presented in Section \ref{sec5}. Some concluding remarks are provided in Section \ref{sec5}.

\section{Mathematical modelling} \label{sec2}
The description as well as the mathematical model of a class of non lower triangular nonlinear systems as follow has been extensively studied in the literature \cite{8908994},  \cite{wang2019backstepping}, \cite{9138416}. The state space dynamic equations describing such class of nonlinear systems are 

\begin{equation}
\begin{fleqn}
\begin{cases}
    d{x}_{i} & =f_{i}^{*}\left(x_{1}, x_{2}, \ldots, x_{n}\right), \quad i=1,2, \ldots, n-1  \tag{1}\\
d{x}_{n} & =g_{n}\left(x_{1}, x_{2}, \ldots, x_{n}\right) u+f_{n}\left(x_{1}, x_{2}, \ldots, x_{n}\right) \\
y & =x_{1}
\end{cases}
\end{fleqn}
\end{equation}

\noindent in the above equation $x_{i} \in R$ are the state variables of the system, $i=1,2, \ldots, n$; $u \in R$ and $y \in R$ are input control signal and the output to the system respectively; $f_{i}^{*}(\cdot)$ and $f_{n}(\cdot)$ are nonlinear functions that are known to be smooth, $i=1,2, \ldots, n ; g_{n}(\cdot)$ is a function being continuous that satisfies $g_{n}(\cdot) \neq 0$.
The objective of the control to be applied in (1) is, to design a controller for the system such that, the output $y$ will track a reference signal given as  $y_{r}$, and therefore all signals in the closed-loop system are evenly distributed and eventually bounded.  

\begin{remark} \label{aus1} 
The system (1) has a non-lower triangular shape, which means that the usual design approaches based on backstepping are not applicable. This study aims to design a switching control approach for the class of systems using an error reduction DSC based methodology.  
\end{remark}

\begin{assumption} \label{aus2} 
    The signs of the term $\partial f_{i}^{*}(\cdot) / \partial x_{i+1}$ are well known, $i=1,2, \ldots, n-1$. For simplicity, it is considered that $\partial f_{i}^{*}(\cdot) / \partial x_{i+1} \geq 0$.
\end{assumption}

\begin{assumption} \label{aus3} 
    The reference signal $y_{r}$ is substantially continuous, and $y_{r}, \dot{y}_{r}$ and $\ddot{y}_{r}$ are sufficiently bounded.
\end{assumption}

\section{Switching backstepping based DSC control design} \label{sec3}
 The objective of our control is to provide a switching control scheme that minimizes errors using backstepping based DSC approaches, ensuring the semi-globally uniformly stable behavior of the closed loop system. Like the conventional backstepping-based DSC approaches, the recursive design process consists of $n$ procedures and two different controllers depending on the magnitude of the error. At each intermediate step $i$, a virtual switching control law $\beta_{i+1}$ is implemented to stabilize the $i$ th subsystem, where $i$ ranges from 1 to $n-1$. In order to avoid cyclic dependence of control laws in the following procedures, a first-order low-pass filter is used to process the virtual control signal $\beta_{i+1}$. The resulting signal from the filter is then fed into the subsequent subsystem. In the last stage, an actual control is formulated to stabilize the $n$th subsystem. The aforementioned process results in the formation of a controller that consists of $n-1$ virtual control laws, $n-1$ first-order filters, and an actual control law. In order to implement the design of the switching control technique, the following theorem are presented.  
 \begin{theorem} \label{aus4} Variable Power Surface Error Function (VPSEF) 
\end{theorem}
Within the framework of backstepping-based dynamic surface control, we present the notion of a Variable Power Surface Error Function (VPSEF), which is denoted as $\psi_i$, that is designed to adjust the influence of surface error on the effectiveness of control, depending on their magnitude. The VPSEF is represented by the following equation:

\begin{equation}
\begin{fleqn}
\begin{cases}
    \psi_1=(x_1-yr)^{p/q}\\ \tag{2} 
     \psi_i=(x_i-a_i)^{p/q}
\end{cases}
\end{fleqn}
\end{equation}

where $x_1$ represent the current state variable, $yr$ denotes the reference signal, the parameters $p$ and $q$ determine the behavior of the function,  with $p>q$ for higher errors and $p<q$ for smaller errors. 

This formulation enables the control response to be adjusted adaptively based on the size of the error. This allows for precise control actions that are customized to different levels of deviation from the reference trajectory. The choice of p and q variables allows for adjusting the VPSEF to the unique dynamics and performance needs of a system. 

To enhance the clarity and conciseness of the design process, system (1) is first reformulated in the following form:

\begin{equation}
\begin{fleqn}
\begin{cases}
    d{x}_{i} & =x_{i+1}+f_{i}\left(x_{1}, x_{2}, \ldots, x_{n}\right), i=1,2, \ldots, n-1  \tag{3}\\
    d{x}_{n} & =g_{n}\left(x_{1}, x_{2}, \ldots, x_{n}\right) u+f_{n}\left(x_{1}, x_{2}, \ldots, x_{n}\right) \\
y & =x_{1}
\end{cases}
\end{fleqn}
\end{equation}

where $f_{i}\left(x_{1}, x_{2}, \ldots, x_{n}\right)=f_{i}^{*}\left(x_{1}, x_{2}, \ldots, x_{n}\right)-x_{i+1}$ are well-known nonlinear smooth function, $i=1,2, \ldots, n-1$.

The detailed methodology for controller design is provided in the following.

Step 1: Given the surface tracking error $\psi_1=x_{1}-y_{r}$ (i.e. the initial surface error). By applying the VPSEF at each iteration step, by checking the switching condition below:
\begin{equation}
\begin{fleqn}
\begin{cases}
      \psi_i=(x_i-y_r)^{p/q}  p>q \\ for- \psi_i > Threshold,  \tag{4} \\
       \psi_i=(x_i-y_r)^{p/q}  p<q  \\ for- \psi_i < Threshold,   \\
       Threshold>0
\end{cases}
\end{fleqn}
\end{equation}

We can obtain the derivative of $\psi_i$ as

\begin{align*}
    \dot\psi_{1} &= \dot{x}_{1}-\dot{y}_{r} \tag{5}\\
     &=x_{2}+f_{1}\left(x_{1}, x_{2}, \ldots, x_{n}\right)-\dot{y}_{r} 
\end{align*}
 If we choose $x_{2}$ as the control signal for subsystem (5), we can design a virtual control that will stabilize the subsystem in the following manner.    
\begin{equation*}
\beta_{2}=-k_{1} \psi_{1}-f_{1}\left(x_{1}, x_{2}, \ldots, x_{n}\right)+\dot{y}_{r} \tag{6}
\end{equation*}
where $k_{1}>0$ being a control parameter and  $\psi_{1}$ is the switching parameter and is obtained as in equation (4). Therefore, we obtained different values of $\beta_{2}$ depending on the magnitude of the errors and thus, dual virtual control.

To address the issue of circular structure, a DSC approach is used to manipulate the virtual control signal. This involves the implementation of a first-order low-pass filter, as shown below.

\begin{equation*}
\sigma_{2} \dot{a}_{2}+a_{2}=\beta_{2} \tag{7}
\end{equation*}  
The variables $\beta_{2}$ and $a_{2}$ represent the input and output of the filter, respectively. Additionally, $\sigma_{2}$ is a positive filtering parameter.

\begin{remark} \label{aus5} 
 The primary focus of the control is to ensure that the tracking error $\psi_{1}$ converges to a limited neighborhood around zero. Hence, if $x_{2}$ is derived as $-k_{1} \psi_{1}-f_{1}\left(x_{1}, x_{2}, \ldots, x_{n}\right)+\dot{y}_{r}$, Subsequently, the subsystem (5) may be mathematically converted into the equation $\dot{\psi}_{1}=-k_{1} \psi_{1}$, where $\psi_{1}$ represents the tracking error. It is important to note that the tracking error $\psi_{1}$ gradually approaches zero in an asymptotic manner. Nevertheless, $x_{2}$ is considered a system state variable and is not subject to design. Thus, a switching virtual control law $\beta_{2}$ is implemented, with the aim of ensuring the convergence of $x_{2}-\beta_{2}$ to neighborhood of zero.   
\end{remark}

\begin{remark} \label{aus6} 
Since the switching virtual control law $\beta_{2}$ incorporates $x_{n}$, the derivative of $\beta_{2}$ will therefore include the real control input $u$. If $\beta_{2}$ is immediately sent to the $x_{2}$-subsystem, the subsequent design of the switching virtual control law $\beta_{3}$ will be depending upon the real control $u$ in the next stage of the design. Consequently, the issue of cyclical reliance of control laws would emerge, making it challenging to practically apply the controller. Hence, the use of DSC methodology is employed to resolve the issue. By allowing the output $a_{2}$ of the filter to reach the following step (step 2), the intended switching virtual control law $\beta_{3}$ will not have a direct dependence on $u$.
\end{remark}

Furthermore, the states $x_{1}, x_{2}, \ldots, x_{n}$ are included in $\beta_{2}$, which implies that the derivative of $\beta_{2}$ will include nonlinear functions $f_{i}\left(x_{1}, x_{2}, \ldots, x_{n}\right)$ for $i=1,2, \ldots, n$. If $\beta_{2}$ is immediately included into the next subsystem, it will result in a complicated design of the switching virtual control rule. The DSC approach is used to send the filter output $a_{2}$ to the next subsystem. The derivative of $a_{2}$ may be derived by the algebraic operation $\dot{a}_{2}=\left(\beta_{2}-a_{2}\right) / \sigma_{2}$. Consequently, the complicated nature of the design is decreased.

Step 2: Let's examine the second error surface, denoted as $\psi_{2}=x_{2}-a_{2}$ by applying the VPSEF as in equation (4). The derivative of $\psi_{2}$ may be calculated as:

\begin{align*}
\dot{\psi}_{2} & =\dot{x}_{2}-\dot{a}_{2} \\
& =x_{3}+f_{2}\left(x_{1}, x_{2}, \ldots, x_{n}\right)-\dot{a}_{2} \tag{8}
\end{align*}
If we choose $x_{3}$ as the control input for subsystem (8), we can create a switching virtual control law to stabilize the subsystem in the following manner:

\begin{equation*}
\beta_{3}=-k_{2} \psi_{2}-f_{2}\left(x_{1}, x_{2}, \ldots, x_{n}\right)+\dot{a}_{2} \tag{9}
\end{equation*}

where $k_{2}>0$ being a control parameter and  $\psi_{2}$ is the switching parameter and is obtained as in equation (4). Therefore, we obtained different values of $\beta_{3}$ depending on the magnitude of the errors and thus, dual virtual control. A first-order low-pass filter is introduced to process $\beta_{3}$ as follows,

\begin{equation*}
\sigma_{3} \dot{a}_{3}+a_{3}=\beta_{3} \tag{10}
\end{equation*}
The filtering parameter is denoted as $\sigma_{3}$ and it must be greater than zero. Similar to step 1, the result of the filter $a_{3}$ will be used in the subsequent stage of the design.

For each step $i$ where $i(i=3, \ldots, n-1)$ : Let's examine the error surface $\psi_{i}$, which is defined as the difference between $x_{i}$ and $a_{i}$. The derivative of $\psi_{i}$ may be calculated as

\begin{align*}
\dot{\psi}_{i} & =\dot{x}_{i}-\dot{a}_{i} \\
& =x_{i+1}+f_{i}\left(x_{1}, x_{2}, \ldots, x_{n}\right)-\dot{a}_{i} \tag{11}
\end{align*}
If we use $x_{i+1}$ as the control input for subsystem (11), we can build a switching virtual control law to stabilize the subsystem in the following manner:

\begin{equation*}
\beta_{i+1}=-k_{i} \psi_{i}-f_{i}\left(x_{1}, x_{2}, \ldots, x_{n}\right)+\dot{a}_{i} \tag{12}
\end{equation*}

where $k_{i}>0$ being a control parameter and  $\psi_{i}$ is the switching parameter and is obtained as in equation (4). Therefore, we obtained different values of $\beta_{i+1}$ depending on the magnitude of the errors and thus, dual virtual control. Implementing a first-order low-pass filter to modify $\beta_{i+1}$ in the following equation:

\begin{equation*}
\sigma_{i+1} \dot{a}_{i+1}+a_{i+1}=\beta_{i+1} \tag{13}
\end{equation*}
The variable $\sigma_{i+1}$ represents a filtering parameter with a positive value. Similar to step 1, the result of the filter $a_{i+1}$ will be used in the subsequent stage of the design.

Step $n$: The final error surface is represented by the equation $\psi_{n}=x_{n}-a_{n}$. The derivative of $\psi_{n}$ may be calculated as

\begin{align*}
\dot{\psi}_{n} & =\dot{x}_{n}-\dot{a}_{n} \\
& =g_{n}\left(x_{1}, x_{2}, \ldots, x_{n}\right) u+f_{n}\left(x_{1}, x_{2}, \ldots, x_{n}\right)-\dot{a}_{n} \tag{14}
\end{align*}
In order to provide stability for subsystem (14) above, an actual switching control is formulated below:
\begin{equation*}
u=\frac{1}{g_{n}\left(x_{1}, x_{2}, \ldots, x_{n}\right)}\left(-k_{n} \psi_{n}-f_{n}\left(x_{1}, x_{2}, \ldots, x_{n}\right)+\dot{a}_{n}\right) \tag{15}
\end{equation*}
where $k_{n}$ is a positive control parameter.

The aforementioned approach yields the designed controller explained by the following

\begin{algorithm}
\caption{Variable Power Surface Error Function backstepping based Dynamic Surface Control Algorithm}
\begin{algorithmic} 
\REQUIRE $Threshold > 0, \psi_{i} $ 
\ENSURE $\psi_{1}=(x_{1}-y_r)$ and $\psi_i=(x_{1}-a_i)$ for $i =2,3,... n-1$
\STATE $\psi_{1,i}\leftarrow Threshold$
\IF{$\psi_{1,i} > Threshold$}
\STATE $\psi_{1,i} \leftarrow (\psi_{1,i})^{p/q} $
\STATE $ p>q $
\ELSE
\STATE $\psi_{1,i} \leftarrow (\psi_1)^{p/q}$
\STATE $p<q$
\ENDIF
\WHILE{$i =1,2,3,... n-1$}
\IF{$\psi_i=\psi_i$ and $p>q$ Lerge error correction }
\STATE $\beta_{2}=-k_{1} \psi_{1}-f_{1}\left(x_{1}, x_{2}, \ldots, x_{n}\right)+\dot{y}_{r}$
\STATE  $\beta_{i+1}=-k_{i} \psi_{i}-f_{i}\left(x_{1}, x_{2}, \ldots, x_{n}\right)+\dot{a}_{i}$
\STATE $u=\frac{1}{g_{n}\left(x_{1}, x_{2}, \ldots, x_{n}\right)}\left(-k_{n} \psi_{n}-f_{n}\left(x_{1}, x_{2}, \ldots, x_{n}\right)+\dot{a}_{n}\right)$
\ELSE[$p<q$ Small error correction]
\STATE $\beta_{2}=-k_{1} \psi_{1}-f_{1}\left(x_{1}, x_{2}, \ldots, x_{n}\right)+\dot{y}_{r}$
\STATE $\beta_{i+1}=-k_{i} \psi_{i}-f_{i}\left(x_{1}, x_{2}, \ldots, x_{n}\right)+\dot{a}_{i}1$
\STATE $u=\frac{1}{g_{n}\left(x_{1}, x_{2}, \ldots, x_{n}\right)}\left(-k_{n} \psi_{n}-f_{n}\left(x_{1}, x_{2}, \ldots, x_{n}\right)+\dot{a}_{n}\right)$
\ENDIF
\ENDWHILE
\end{algorithmic}
\end{algorithm}

\section{Stability Analysis} \label{sec4}
Given the nonlinear system in (1), with the controller designed as in equation (15). If we consider the error of the dynamics corresponding to the final state of the system to be:
\begin{align*}
     \psi_n=(x_n-a_n)^{p/q}\tag{17}
\end{align*}

Where $a_n$ is the desired trajectory for $x_n$.

let the error be,  $e=x_n-a_n$, where the error dynamics is given as:

\begin{align*}
    \dot{x}_{n}-\dot{a}_{n}=-k_n(x_n-a_n)
\end{align*}
 and this simplifies to:
 
 \begin{align*}
    \dot{e}=-k_{n} e
\end{align*}
The given equation is a linear differential equation with a solution
\begin{align*}
    e(t)=e(0)e^{-k_{n}t}
\end{align*}
As $ t \rightarrow \infty$ $e(t) \rightarrow 0 $ if $k_{n} > 0$, this shows that, $ 
x_{n} \rightarrow a_{n}$. Since, $x_{n}$ converges to $a_{n}$ it is important that we understand the impact of this on the whole system, particularly $x_{n-1}$. 

Consider the dynamics for the equation of $x_{n-1}$ as, 

\begin{align*}
    \dot{x}_{n-1} & =f^*_{n-1}\left(x_{1}, x_{2}, \ldots, x_{n}\right)
\end{align*}

As $ x_{n} \rightarrow a_n$ we will substitute $x_n=a_n$ into $f^*_{n-1}$

\begin{align*}
    \dot{x}_{n-1} & =f^*_{n-1}\left(x_{1}, x_{2}, \ldots, a_{n}\right)
\end{align*}

$f^*_{n-1}$ is designed such that if $ x_{n} = a_n$, $ x_{n-1} \rightarrow a_{n-1}$ and will continue throughout the system. Therefore, It is possible to show recursively that each state $x_i$ converges towards its desired trajectory $a_i$ for $i=n-1, n-2,..., 1$

Overall, the switching VPSEF  backstepping based DSC approach developed in the previous section ensures the stability of each subsystem step by step as the controller ensures that $ x_{n} \rightarrow a_n$. Furthermore, in the $n-1$ stage, given that $ x_{n} \rightarrow a_n$, the designed controller ensures that $a_{n-1}$ is designed such that $ x_{n-1} \rightarrow a_{n-1}$. Finally, $x_i$ is assured to converge to $a_i$, thus, stabilizing the entire system.

\section{Simulation results} \label{sec5}
In this section, We present the aforementioned theoretical result, to validate the efficacy of the proposed VPSEF Back-stepping Based DSC Control method. Let us consider a nonlinear system that is non-lower triangular and has a third-order as shown below.

\begin{align*}
      \dot{x}_{1} & =x_{1}^{2}+x_{2}^{3}+x_{3} \\
\dot{x}_{2} & =x_{1}^{2} x_{2}+x_{3}^{5} \\
\dot{x}_{3} & =u+x_{1} x_{2} x_{3}^{2} \\
y & =x_{1} 
\end{align*}
    
The control goal for the system above is to make $x(t)$ to track $y_r=sin(t)$. 
Following Algorithm 1, a VPSEF Back-stepping Based DSC Control is designed with threshold of 0.1. Based on the control design approach outlined in Algorithm 1, the controller for the system may be created as follows:

  \begin{equation}
\begin{fleqn}
\begin{cases}
     \beta_{2} & =-k_{1}(\psi_1)-\left(x_{1}^{2}+x_{2}^{3}+x_{3}-x_{2}\right)+\cos t  \tag{16}\\
\beta_{3} & =-k_{2}(\psi_2)-\left(x_{1}^{2} x_{2}+x_{3}^{5}-x_{3}\right)+\dot{a}_{2} \\
u & =-k_{3}(\psi_3)-x_{1} x_{2} x_{3}^{2}+\dot{a}_{3}
\end{cases}
\end{fleqn}
\end{equation}   

where the variables $a_{2}$ and $a_{3}$ represent the outputs of first-order filters. $\beta_{2}=\sigma_{2} \dot{a}_{2}+$ $a_{2}$ and $\beta_{3}=\sigma_{3} \dot{a}_{3}+a_{3}$ respectively. The selected control parameters for simulation are $k_{1}=3$, $k_{2}=3$, and $k_{3}=3$. The initial states of the system are $x_{1}(0)=0, x_{2}(0)=-1, x_{3}(0)=1$. The initial states of the filters are $a_{2}(0)=0, a_{3}(0)=0$.

In the process of implementing $\beta_{3}, \dot{a}_{2}$ is obtained by $\dot{a}_{2}=\left(\beta_{2}-a_{2}\right) / \sigma_{2}$. In order to prevent an excessively large signal of $\beta_{3}$ during the early phase of operation of the control system, the filtering parameter $\sigma_{2}$ is selected as $\sigma_{2}=\exp (-t)+0.05$. By using this approach, the output of the filter may closely track the input of the filter over time, ensuring that the value of $\dot{a}_{2}$ remains within an acceptable range. We use a similar strategy for the implementation of $u$, and the selection of the filtering parameter is determined as $\sigma_{3}=\exp (-t)+0.05$. 

The figures below illustrate the simulation result of this particular scenario. It is evident that all signals exhibit uniform ultimate boundedness. Figure 1 demonstrates the system's ability to rapidly and accurately track the reference signal $y_{r}$ with the system output $y$. The time required for adjustment is less than 1 second, and the maximum steady state tracking error is less than $1 \%$. This is due to the error correction mechanism, where the initial error magnitudes are being reduced preventing the system from initial overshoot. The control input signal and the system states are shown in figure 2 and 3 respectively. Figure 5 and 6 depict the input and output signals of the two first-order low-pass filters respectively. Figure 4 clearly shows the impact of the proposed technique, where by the tracking error is less than $1 \%$. The stability of the system and the achievement of perfect tracking performance has being obtained using the suggested controller designed in this work. Figure 7 depict the high error reduction in traccking the reference $y_r$ resulting from the proposed control algorithm compared to the control technique presented in \cite{wang2019backstepping}.

Likewise, when the reference signal is taken as $y_{r}=1-e^{-t}$. To further validate the effectiveness of the suggested control mechanism given in (16) another reference signal $y_{r}=1-e^{-t}$ and the simulation result is given in the figures 8 to 13. The effective achievement of stability and tracking performance in a closed-loop system in question is clearly seen. Figure $7-12$ shows the system output, control input, system states, tracking error and the first order filter respectively related to the second reference using the proposed control design.  
\begin{figure}
\centering
\includegraphics[width=1\linewidth]{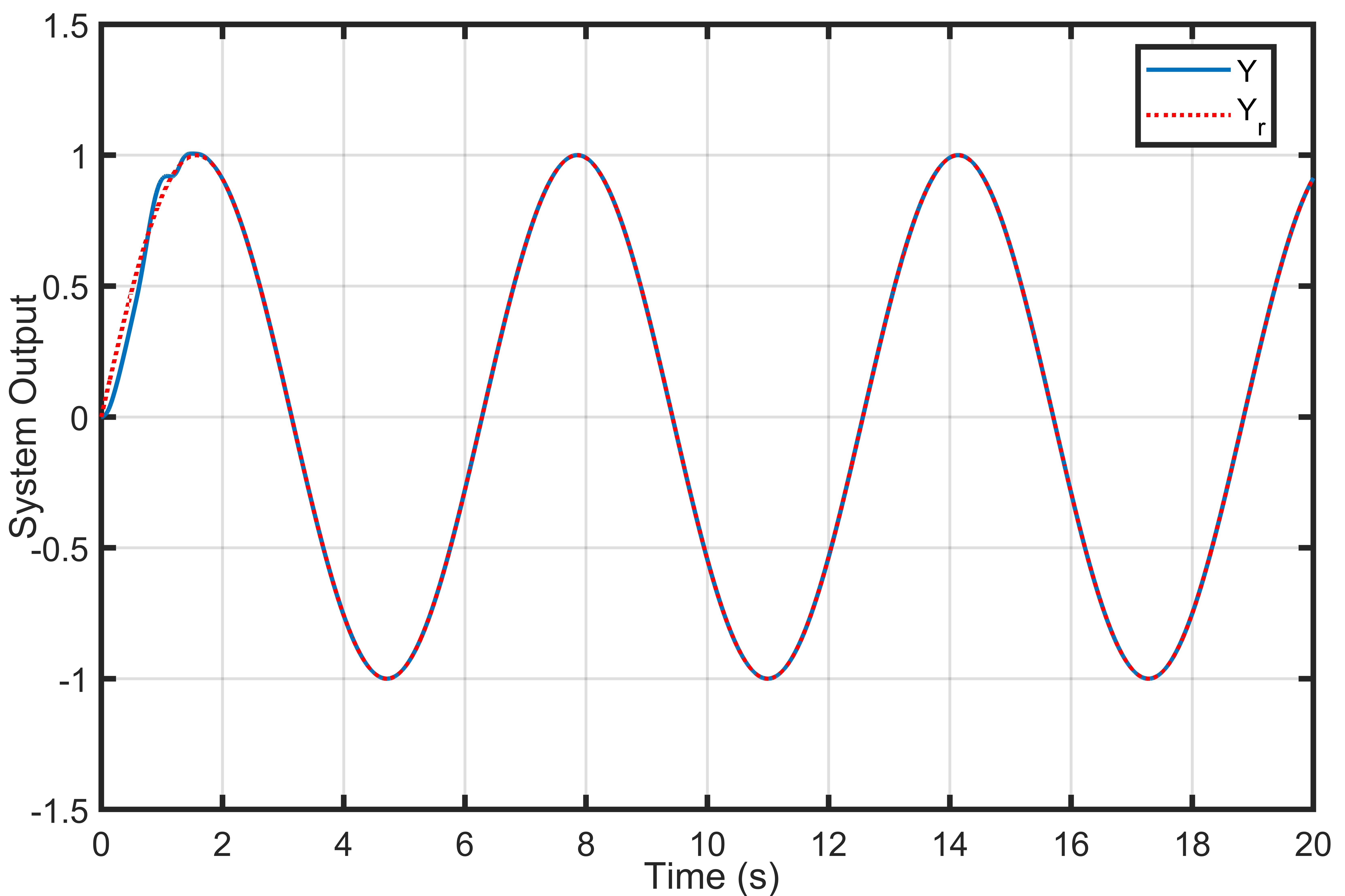}
\caption{System Output }
 \label{Fig:three1}
\end{figure}

\begin{figure}
\centering
\includegraphics[width=1\linewidth]{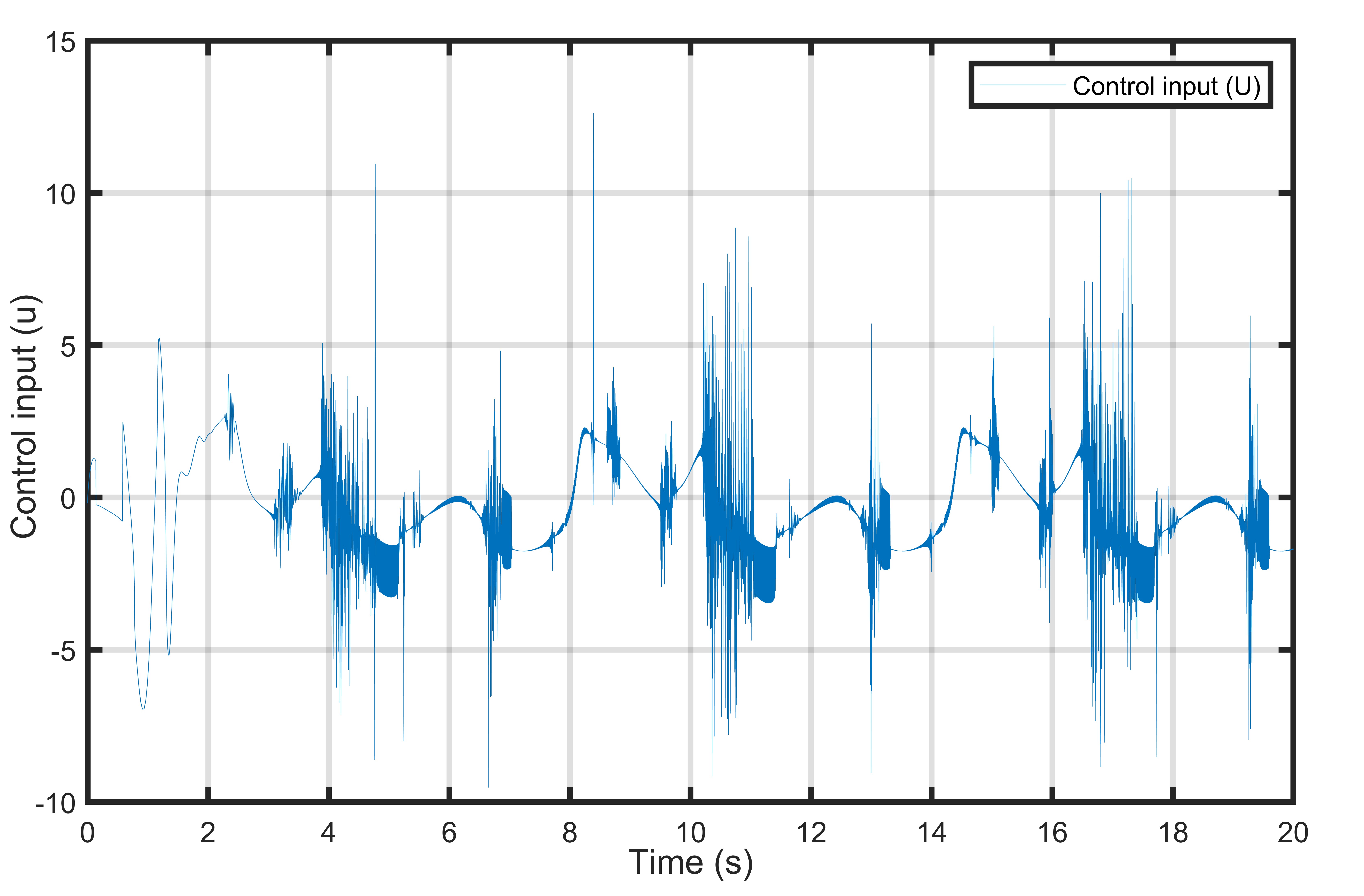}
\caption{Control Input}
 \label{Fig:three2}
\end{figure}

\begin{figure}
\centering
\includegraphics[width=1\linewidth]{ 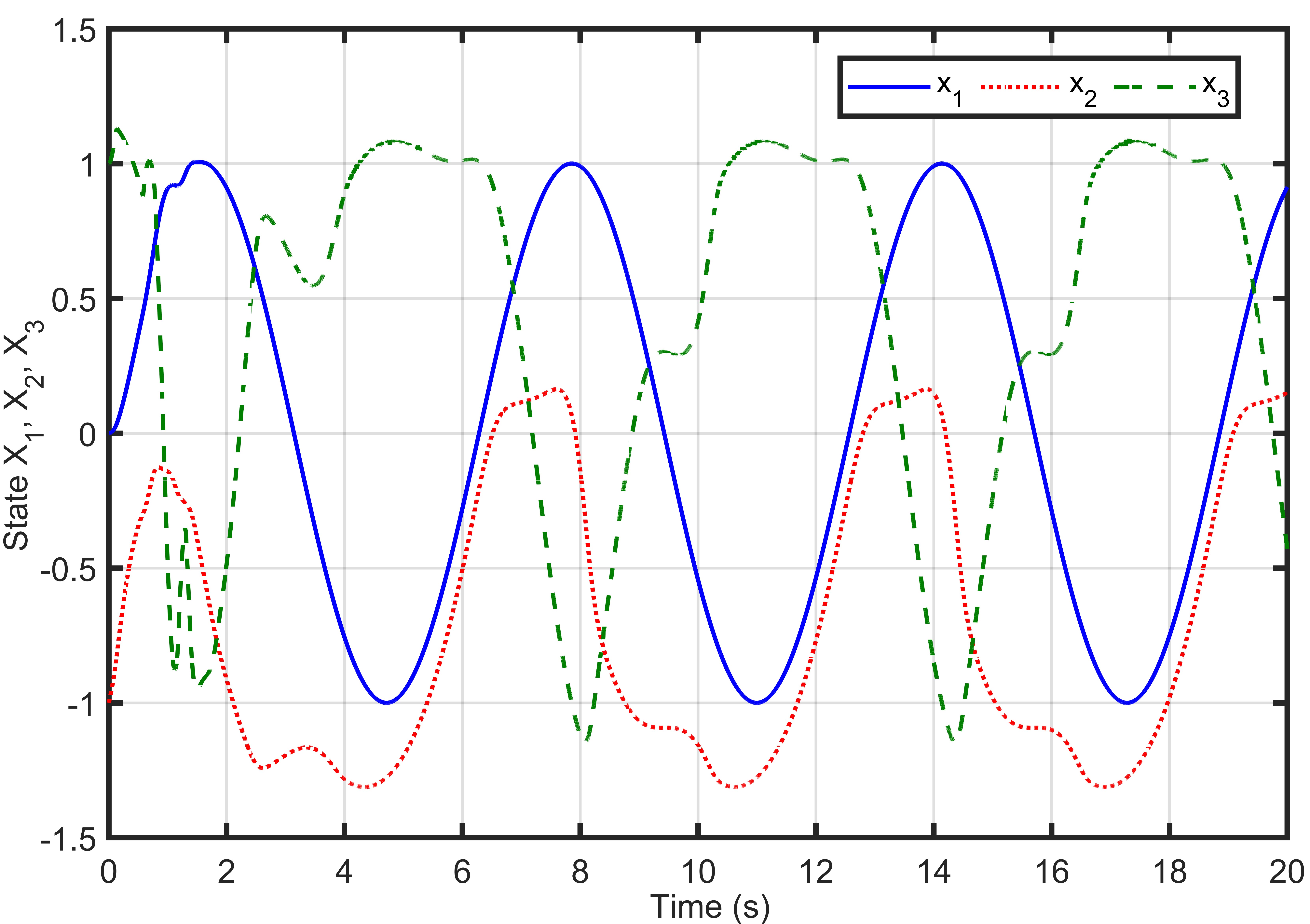}
\caption{System States }
 \label{Fig:three3}
\end{figure}

\begin{figure}
\centering
\includegraphics[width=1\linewidth]{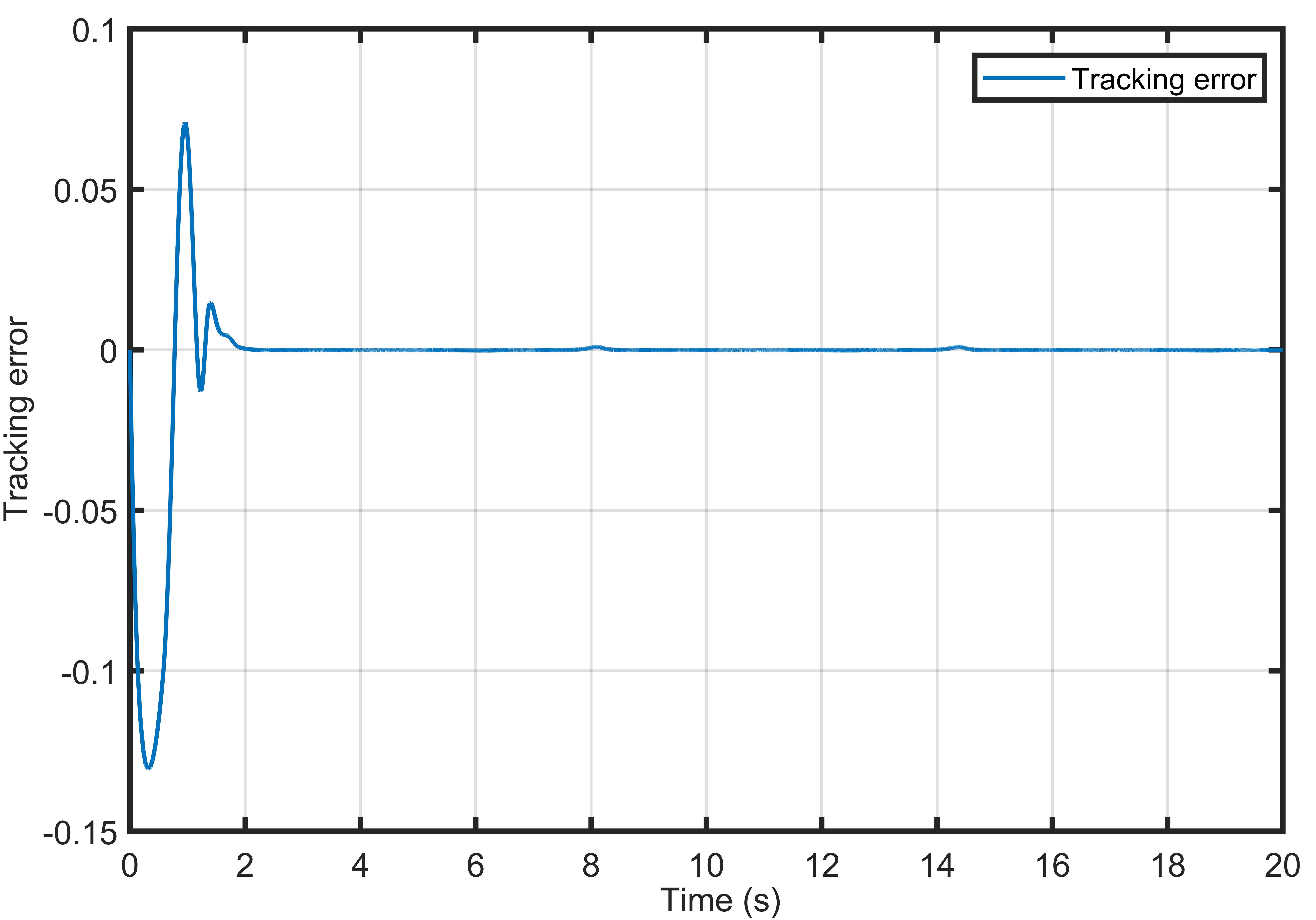}
\caption{Tracking error }
 \label{Fig:three4}
\end{figure}

\begin{figure}
\centering
\includegraphics[width=1\linewidth]{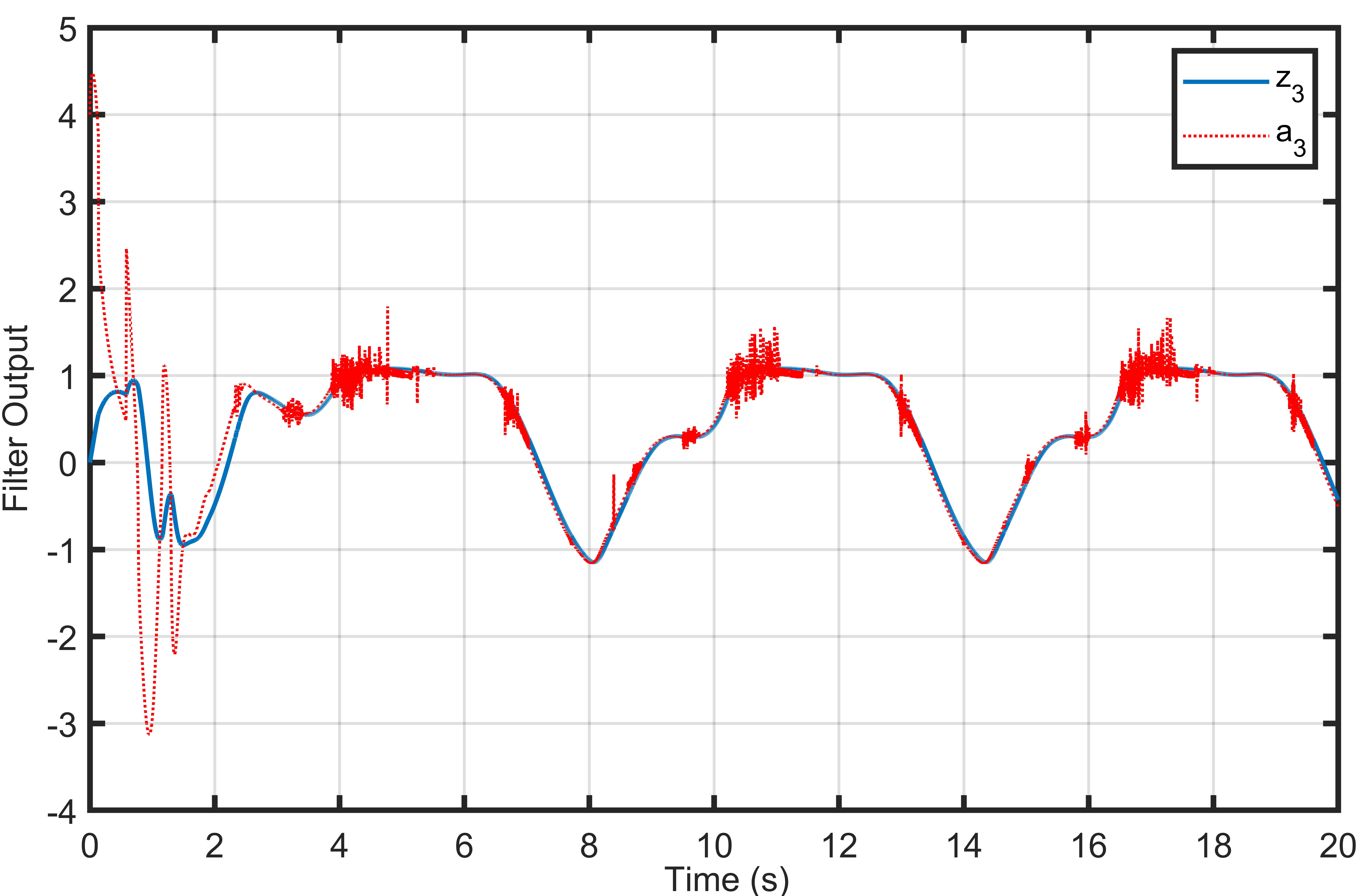}
\caption{First Order Filter }
 \label{Fig:three5}
\end{figure}

\begin{figure}
\centering
\includegraphics[width=1\linewidth]{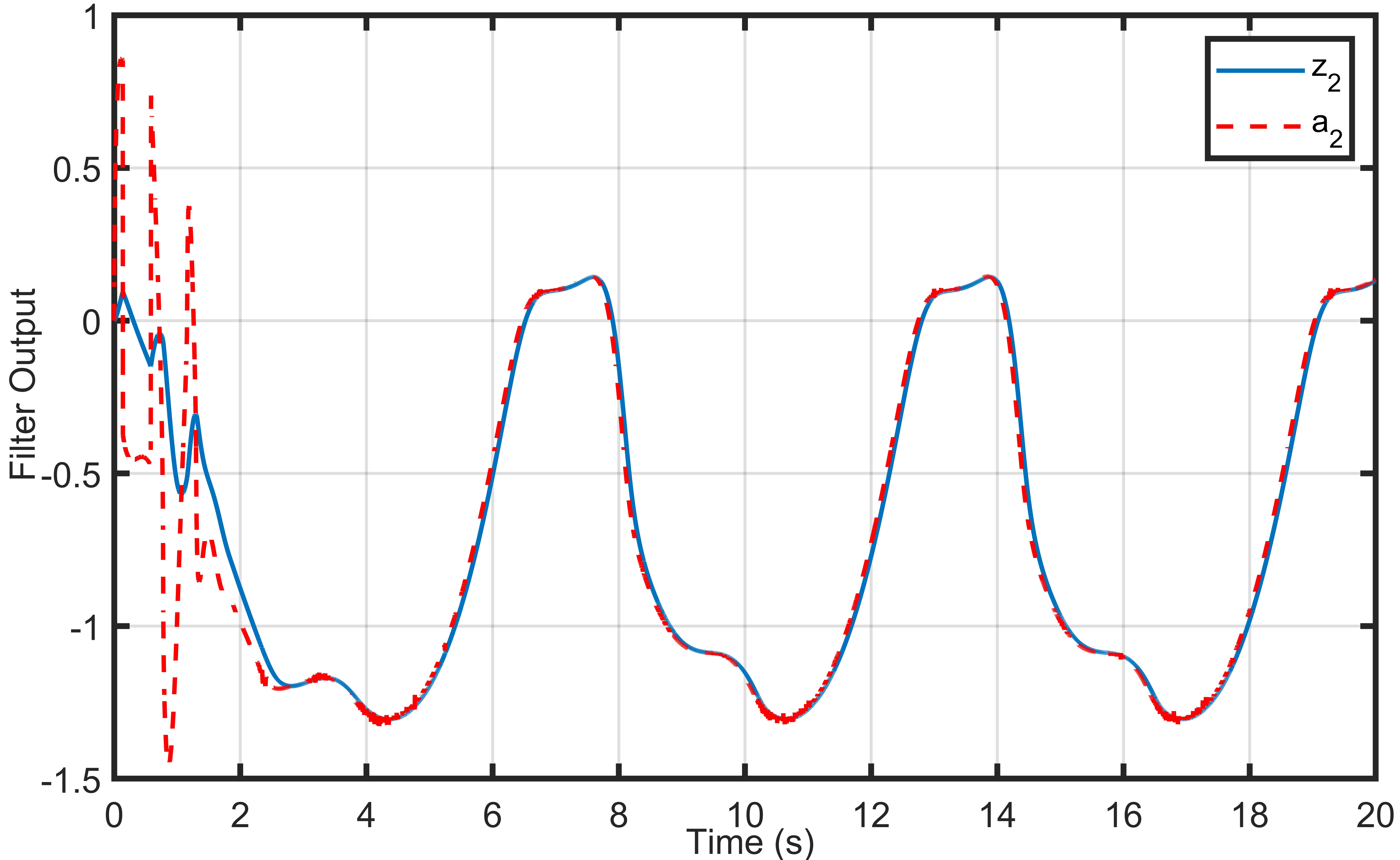}
\caption{First Order Filter}
 \label{Fig:three6}
\end{figure}

\begin{figure}
\centering
\includegraphics[width=1\linewidth]{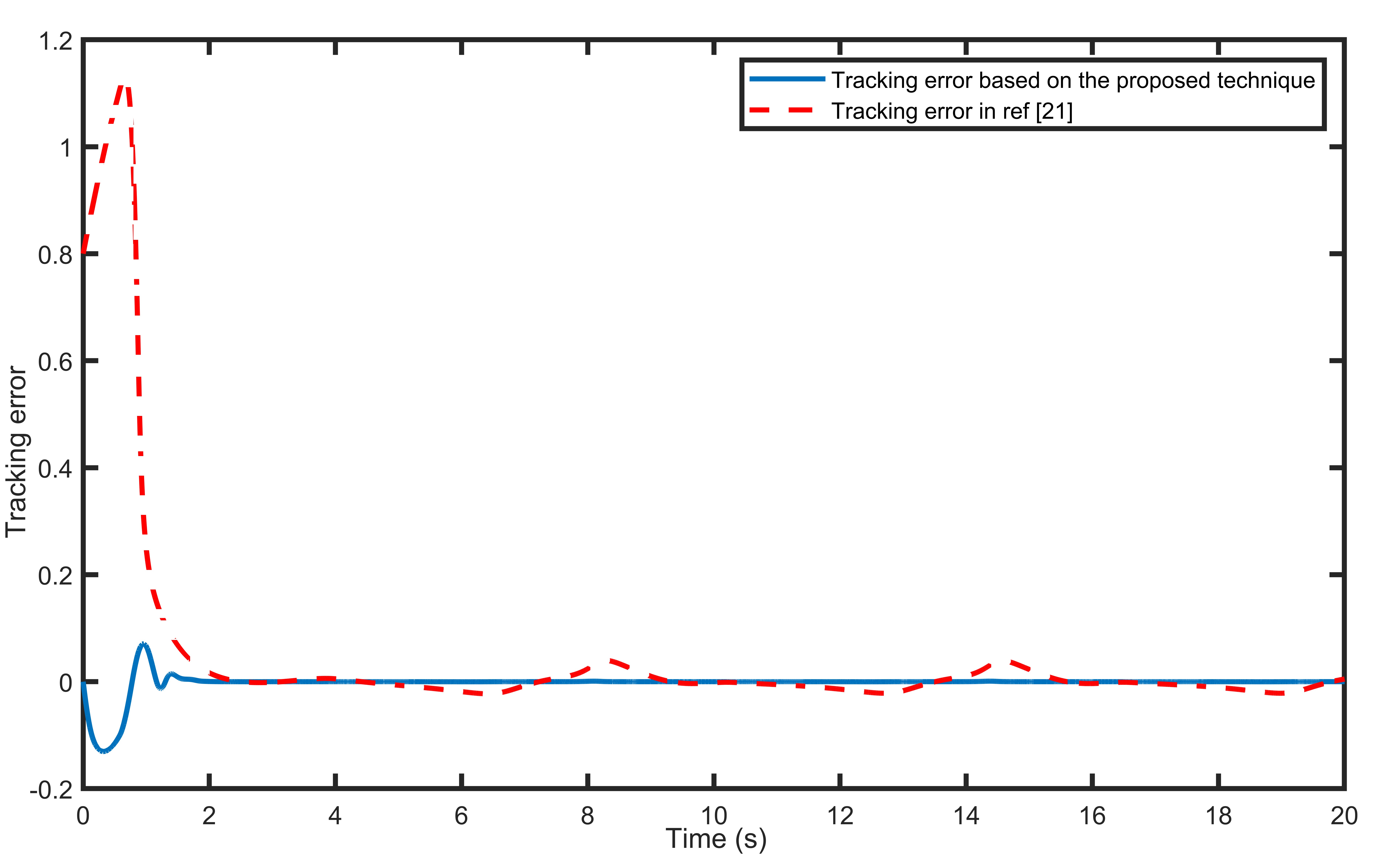}
\caption{Error comparison with ref [21] }
 \label{Fig:three7}
\end{figure}

\begin{figure}
\centering
\includegraphics[width=1\linewidth]{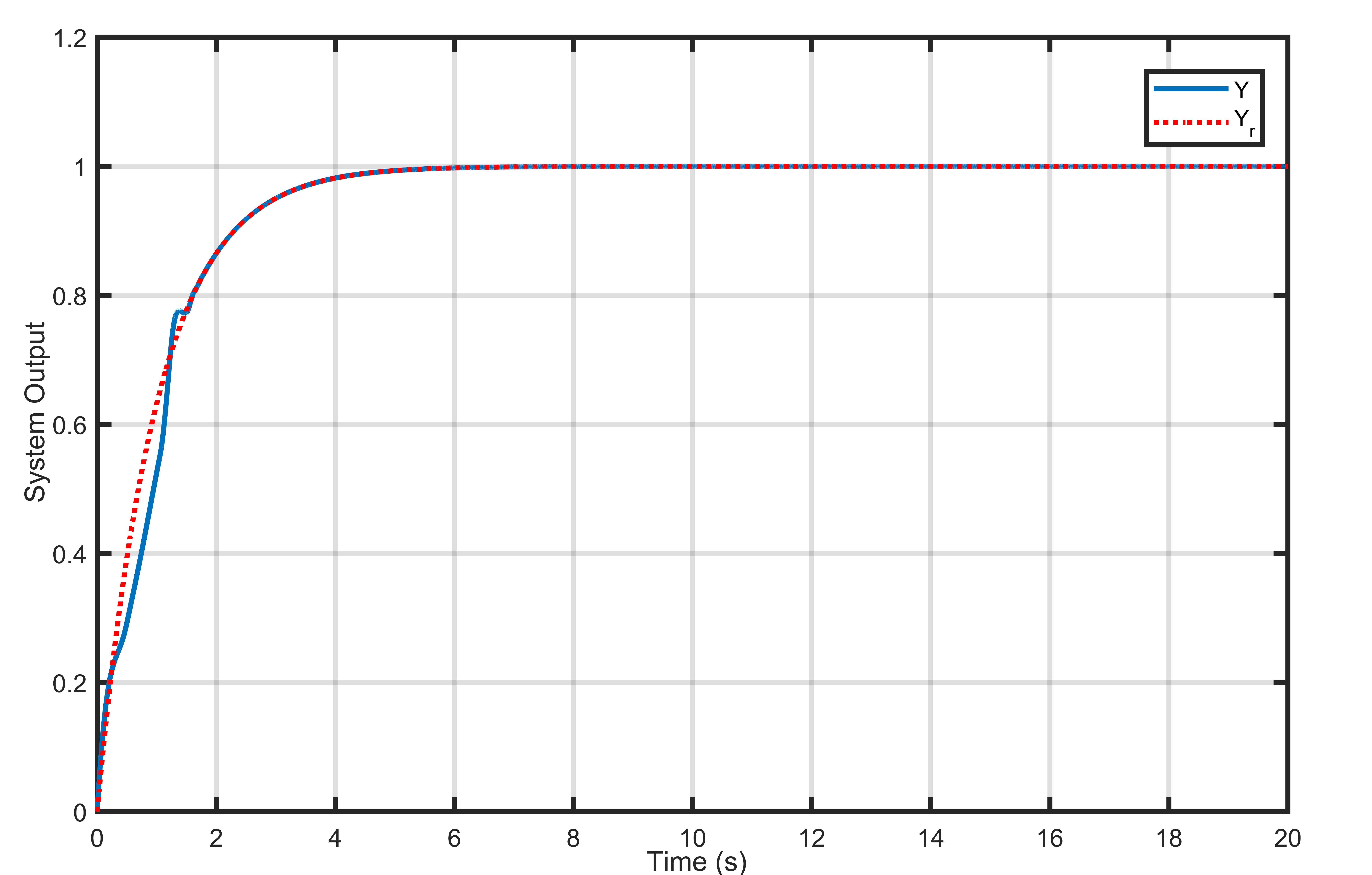}
\caption{System Output}
 \label{Fig:three8}
\end{figure}

\begin{figure}
\centering
\includegraphics[width=1\linewidth]{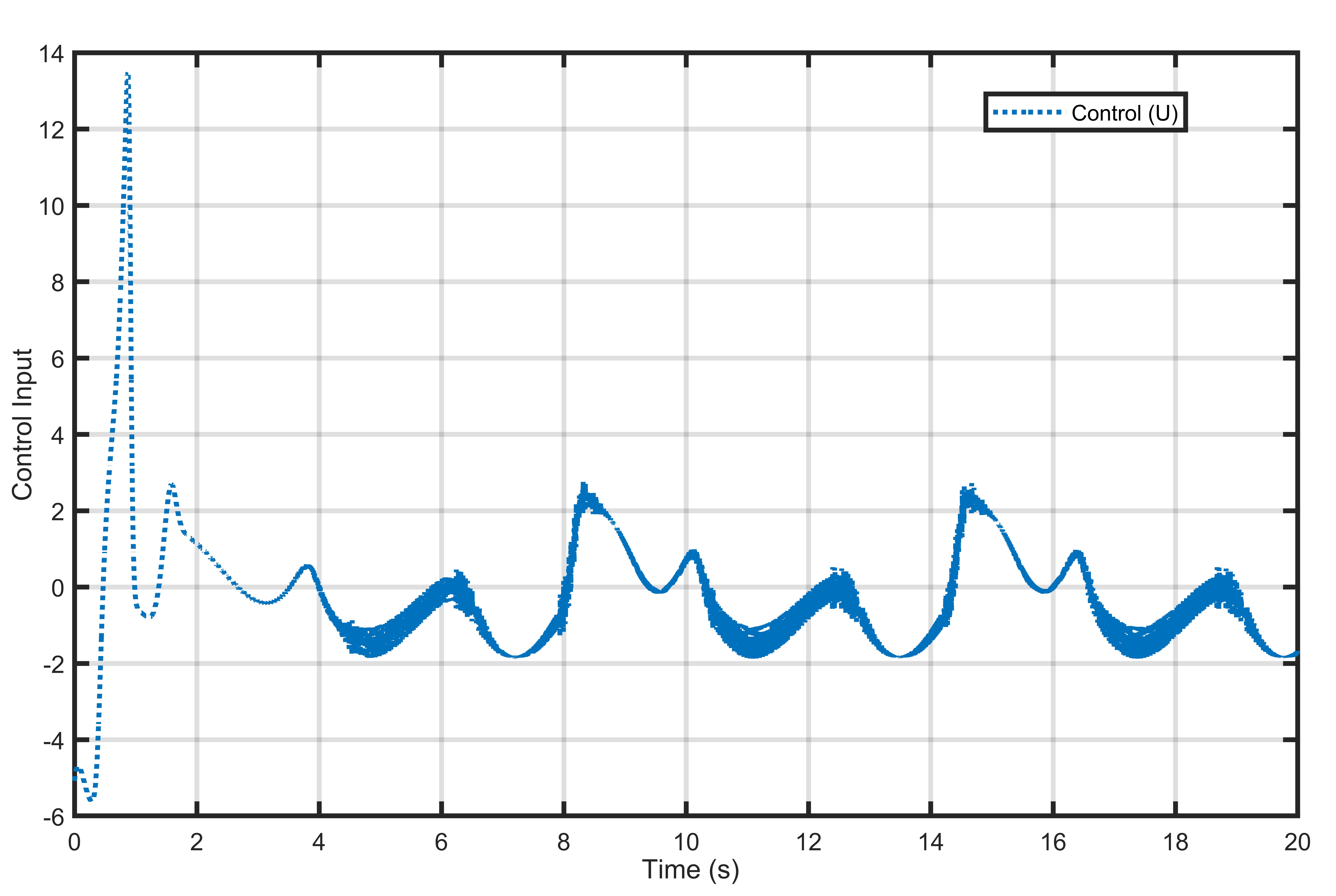}
\caption{Control Input}
 \label{Fig:three9}
\end{figure}

\begin{figure}
\centering
\includegraphics[width=1\linewidth]{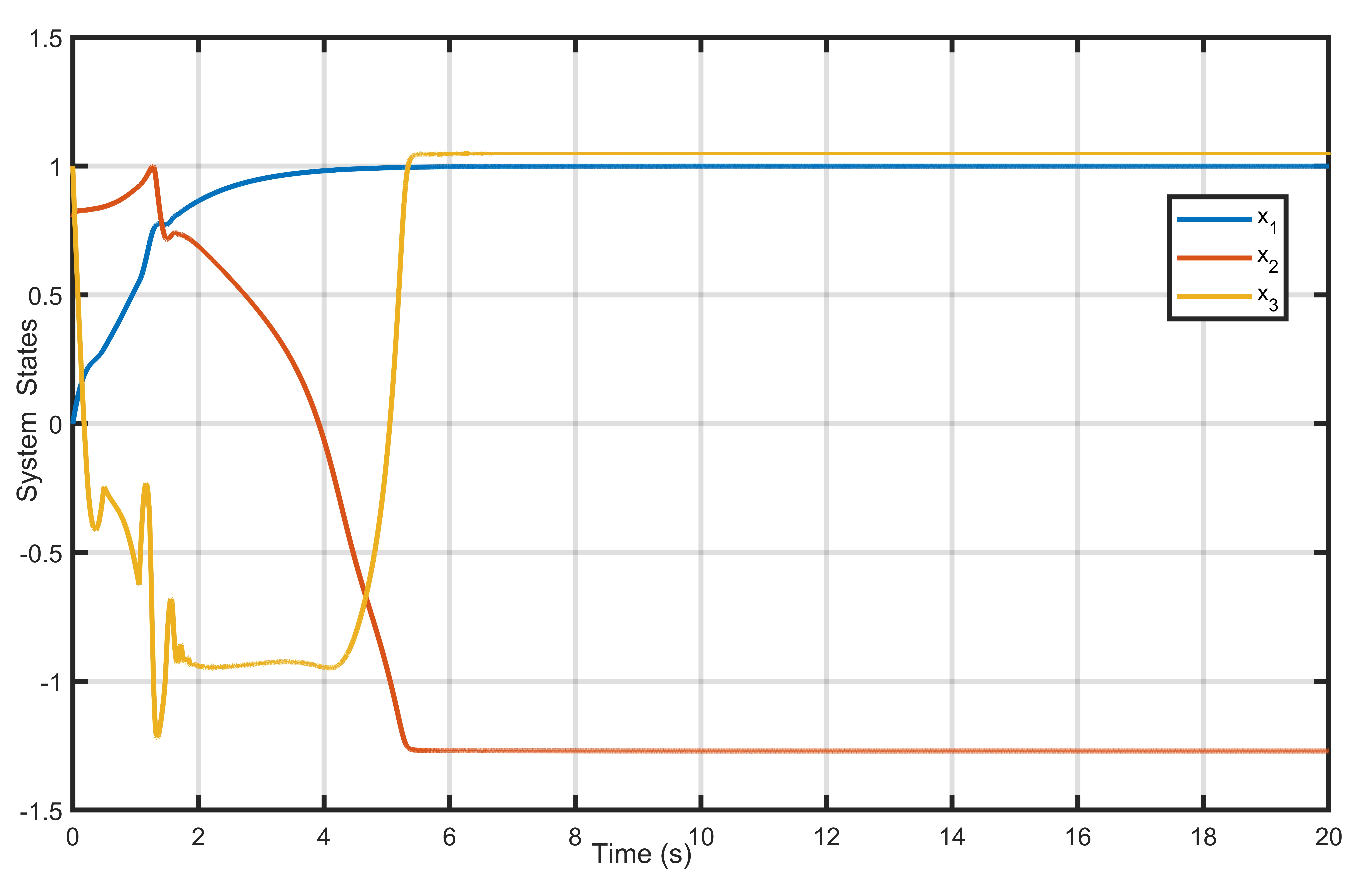}
\caption{System States}
 \label{Fig:three10}
\end{figure}

\begin{figure}
\centering
\includegraphics[width=1\linewidth]{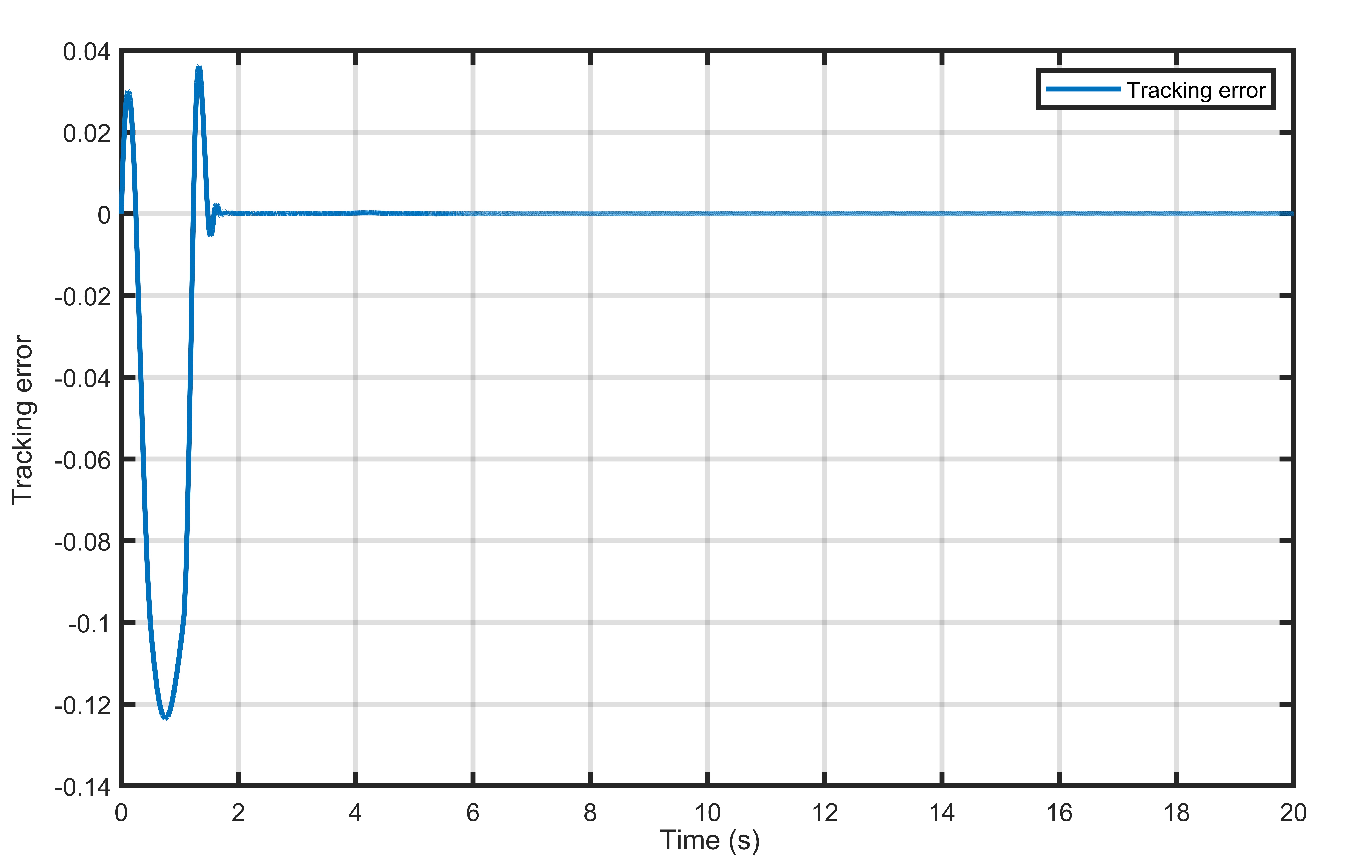}
\caption{Tracking Error}
 \label{Fig:three11}
\end{figure}

\begin{figure}
\centering
\includegraphics[width=1\linewidth]{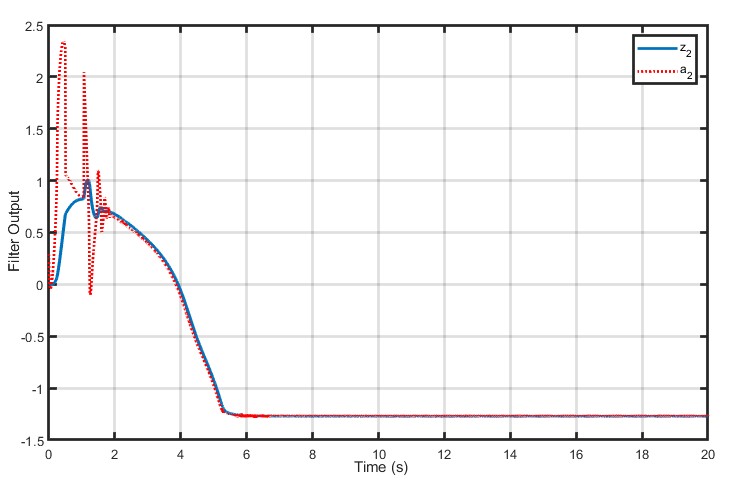}
\caption{First Order Filter}
 \label{Fig:three12}
\end{figure}

\begin{figure}
\centering
\includegraphics[width=1\linewidth]{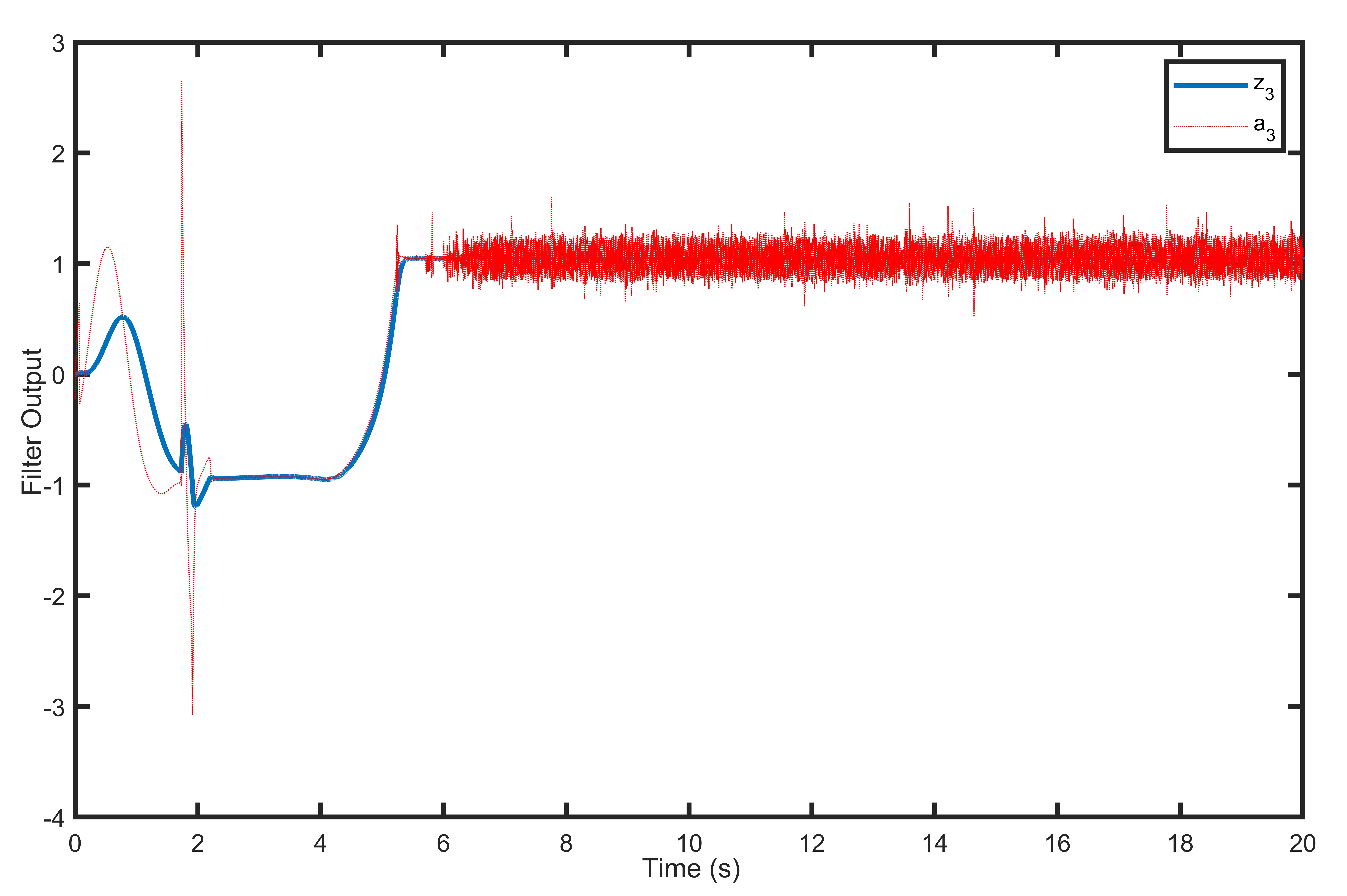}
\caption{First Order Filter}
 \label{Fig:three13}
\end{figure}

\section{Conclusions} \label{sec6}
This article introduces a new technique called Variable Power Surface Error Function backstepping based Dynamic Surface Control for non lower-triangular non linear systems. The approach is supported by strong theoretical proofs. The control law is switching based on the magnitude of error in order to mitigate the high impact of these errors. The designed control technique not only ensures accurate and prescribed settling time tracking of the reference, but also actively reduces oscillations in the tracking error and intermediate errors. Furthermore, the use of the DCS in this approach effectively eliminates the issue associated with the circular design. The stability analysis shows that all signals in the closed-loop control system are uniformly ultimately bounded, and the tracking error converges to a small vicinity of zero. Two simulations were conducted using various reference signals to illustrate the enhanced trajectory tracking achieved by the suggested technique. The simulations showed a reduction in tracking error and changes in reference signals, highlighting the effectiveness of the proposed approach.




\textbf{Declarations}

\textbf{    Ethical Approval}

Not applicable

\textbf{    Funding}

No funding was received for this work.

\bibliographystyle{elsarticle-num-names}
\bibliography{Main.bib}







\end{document}